\documentclass[]{article}

\usepackage{graphicx}
\usepackage[capposition=top]{floatrow}
\usepackage[margin=0.8in]{geometry}
\usepackage{amsmath}
\usepackage{wrapfig}
\usepackage{bm}
\usepackage{floatflt}
\usepackage{hyperref}
\usepackage{multicol}
\usepackage{lscape}
\usepackage{pdfpages}
\title{Delicatessen: Automated Estimating Equations in Python}
\author{Paul N Zivich\textsuperscript{1,2}, Mark Klose\textsuperscript{1}, Stephen R Cole\textsuperscript{1}, Jessie K Edwards\textsuperscript{1}, Bonnie E Shook-Sa\textsuperscript{3}}
\date{%
	\footnotesize
	\textsuperscript{1}Department of Epidemiology, Gillings School of Global Public Health, University of North Carolina at Chapel Hill, Chapel Hill, NC\\%
	\textsuperscript{2}Institute of Global Health and Infectious Diseases, School of Medicine, University of North Carolina at Chapel Hill, Chapel Hill, NC\\%
	\textsuperscript{3}Department of Biostatistics, Gillings School of Global Public Health, University of North Carolina at Chapel Hill, Chapel Hill, NC\\[2ex]%
	\today
}

\begin{document}
	
\maketitle

\noindent\rule{1.0\linewidth}{0.4pt}

\begin{abstract}
	Estimating equation theory provides a unified framework for statistical modeling and inference. Applying estimating equations has historically involved extensive matrix algebra and by-hand differentiation of complex functions. Here, we introduce \texttt{delicatessen}, a Python library that automates those tedious calculations, lowering the barrier to adoption in quantitative biological and life science research. To highlight the utility of \texttt{delicatessen} for the acceleration of scientific research, we provide illustrative examples of linear regression with outliers, estimation of a dose-response curve, and standardization of the mean. Across these and other settings, \texttt{delicatessen} streamlines transparent and reproducible advanced statistical procedures for modern data analysis.
\end{abstract}

\noindent\rule{1.0\linewidth}{0.4pt}
~\\

Estimating equation theory is a general large-sample statistical framework that encapsulates many routinely used quantitative methods \cite{Stefanski2002, Ross2024}.  An estimating equation is defined as $\sum_{i=1}^{n} \psi(\mathbf{Z}_i; \boldsymbol{\theta}) = 0$, where $\boldsymbol{\theta} = \{\theta_1, \theta_2, ..., \theta_v\}$ is the parameter vector, $\psi(.)$ is a estimating function that returns a vector of dimension $v$, and $\mathbf{Z}_i$ are the observed data for $n$ units. The estimator, sometimes referred to as M-estimator or Z-estimator, is then the solution to the estimating equation, $\hat{\boldsymbol{\theta}}$. The variance of $\hat{\boldsymbol{\theta}}$ can be estimated via the empirical sandwich variance estimator (see Appendix), which remains consistent even when parameters in $\boldsymbol{\theta}$ depend on each other in complex ways.

Despite their extraordinary flexibility and potential to accelerate life science research, widespread adoption of estimating equations is hindered by tedious matrix calculations and differentiation. To address this barrier, we developed \texttt{delicatessen}, an open-source Python library that automates the underlying computations of estimating equations. Unlike existing statistical software for estimating equations \cite{Saul2020, Pedregosa2011, Seabold2010}, \texttt{delicatessen} provides support for both arbitrary estimating functions and an extensive set of pre-built estimating equations widespread in scientific research (Appendix Table 1). Built-in estimating equations include generalized linear models, generalized additive models, survival analysis models, causal inference methods, instrumental variables, measurement error correction, and more. These built-in functions can be easily combined with user-specified estimating equations. There is further support for finite-sample corrections and clustered observations. Currently, the online documentation for \texttt{delicatessen} hosts more than 30 open-source replications of analyses in biomedical, pharmacology, chemistry, astronomy, economics, and other areas of research. To highlight the use of \texttt{delicatessen} for life science research, we showcase three applications: linear regression with outliers, dose-response models, and standardization. 

Within \texttt{delicatessen}, the estimator for a user-specified estimating function is evaluated through the \texttt{MEstimator} class, which solves for $\hat{\boldsymbol{\theta}}$ using root-finding algorithms implemented in \texttt{SciPy} \cite{Virtanen2020} or a user-provided root-finding algorithm. After solving for $\hat{\boldsymbol{\theta}}$, the empirical sandwich covariance matrix is constructed from $\psi(\cdot)$. This step involves matrix algebra operations implemented in \texttt{NumPy} \cite{Harris2020}, and evaluating the partial derivatives of $\psi(\cdot)$ using either numerical approximation from \texttt{SciPy} or a dedicated implementation of automatic differentiation. 

To begin, consider linear regression with outliers \cite{Altman2015, Altman2016}. A common approach to handling outliers is to exclude them. However, exclusion should only be done when outliers are unambiguously a result of experimental error \cite{Altman2016}. Yet, including outliers with simple linear regression can lead to unstable estimates. Robust regression is an alternative, whereby outliers contribute but their influence is curtailed \cite{Huber2004}. The estimating functions for the intercept and slope with robust linear regression are
\[\psi(Y_i, X_i, \boldsymbol{\alpha}) =
\begin{bmatrix}
	f_k(Y_i - (1, X_i) \boldsymbol{\alpha}^T)1 \\
	f_k(Y_i - (1, X_i) \boldsymbol{\alpha}^T)X_i
\end{bmatrix}
\]
where $Y_i$ is the dependent variable, $X_i$ is the independent variable, and $\boldsymbol{\alpha}=(\alpha_0,\alpha_1)$ is the parameter vector for the regression model. Here, 
\[ f_k(z) = 
\begin{cases}
	-k & \text{if } -k < z \\
	z & \text{if } -k \le z \le k \\
	k & \text{if } k < z
\end{cases}
\] 
\setlength{\columnsep}{20pt}%
\begin{wrapfigure}{l}{0.45\linewidth}
	\centering
	\caption{Linear regression (a), dose-response curve (b), and standardized mean (c)}
	\includegraphics[width=1.0\linewidth]{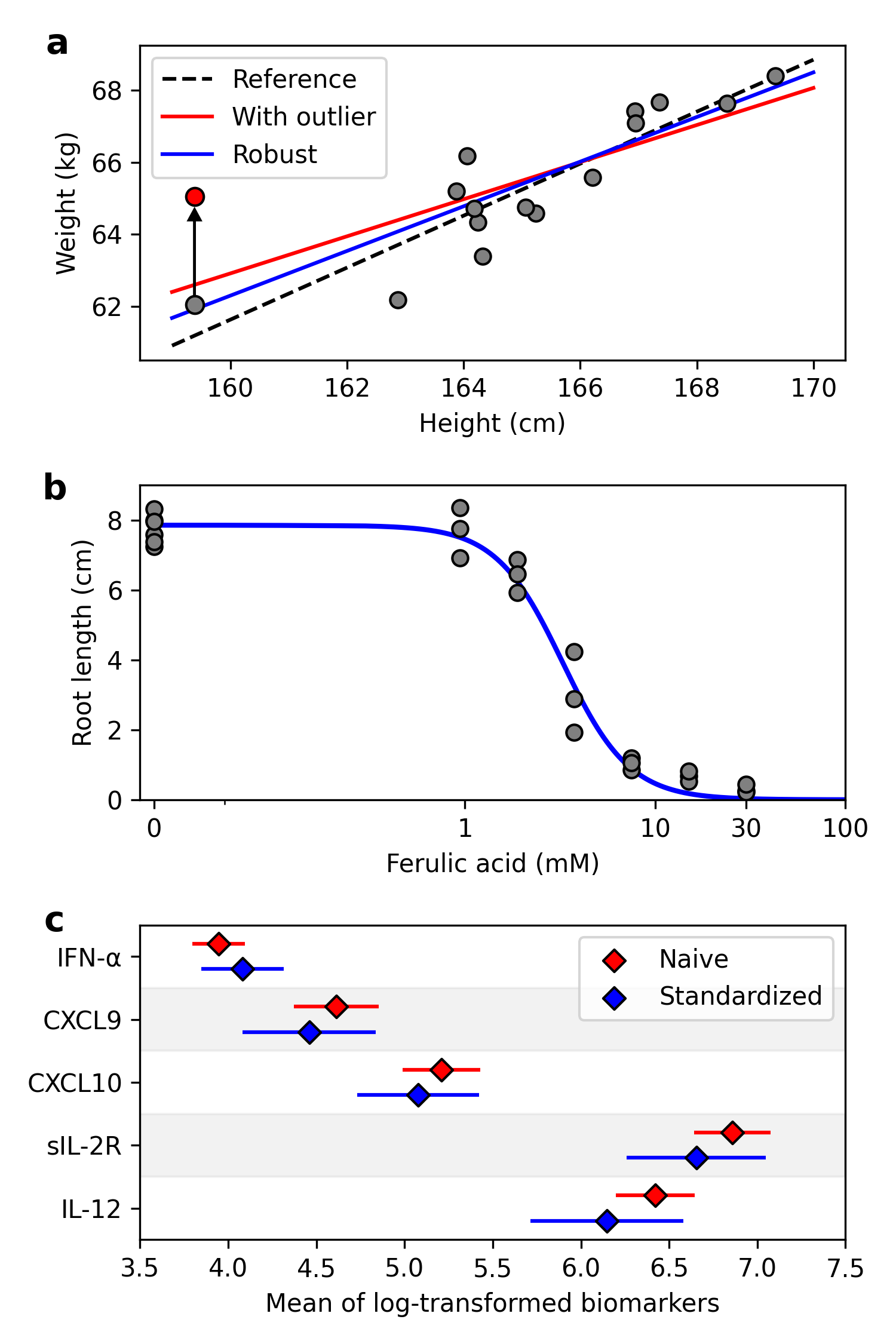}
	\floatfoot{
		\textbf{a}: The reference line was estimated using simple linear regression based on the gray data points. The `with outlier' simple and `robust' lines were estimated with the induced outlier, indicated by the red data point.\\
		\textbf{b}: Dose-response curve for herbicide on ryegrass root length. Dots indicate the data points, and the blue line is the estimated dose-response curve.\\
		\textbf{c}: Means of the log-transformed biomarkers. Naive corresponds to the means for the original $n=57$ HIV patients. Standardized corresponds to the inverse odds weighted means. Lines indicate the 95\% confidence intervals.}
\end{wrapfigure}
\noindent
which limits the maximum influence of outliers. To illustrate, 15 observations were simulated following Altman \& Krzywinski (2016) \cite{Altman2016}. All models were fit using \texttt{delicatessen} with the built-in estimating equations for robust regression (i.e., \texttt{ee\_robust\_regression}). For reference, a linear model was fit to the simulated data (Figure 1a, slope: $0.72$). An outlying value of $Y_i$ was then generated by replacing $Y_i$ with $Y_i + 3$ for the observation with the smallest $X_i$ value. When fitting a linear model with this outlier, the slope is noticeably decreased ($0.52$). With robust regression and $k=1.345$, the influence of the outlier was curtailed and the estimated slope ($0.62$) was closer to the reference.

As a second example, consider estimating the dose-response relationship between a herbicide on ryegrass root length ($n=24$) \cite{Inderjit2002, Ritz2015}. Since there is a natural lower bound of zero on root length, a three-parameter log-logistic (3PL) model was used. Parameters of the 3PL model are the halfway maximum effective concentration ($\gamma_1$), steepness of the dose-response curve ($\gamma_2$), and upper limit on the response ($\gamma_3$). Let $\psi_{PL}(D_i, R_i; \boldsymbol{\gamma})$ denote the corresponding estimating function (see Appendix), where $D_i$ is the dose of ferulic acid (mM) and $R_i$ is root length (cm). Concentration levels other than the halfway maximum effective concentration may also be of interest. Suppose interest was in the 20\% effective concentration ($\delta_{0.2}$). While it is straightforward to estimate $\delta_{0.2}$ from $\hat{\boldsymbol{\gamma}}$, the dependence of the variance of $\hat{\delta_{0.2}}$ on the variance of $\hat{\boldsymbol{\gamma}}$ complicates statistical inference. Methods to appropriately estimate the variance of $\hat{\delta}_{0.2}$ can be burdensome to apply manually (i.e., delta-method) or are computationally burdensome (e.g., bootstrapping \cite{Kulesa2015}, Monte Carlo methods \cite{Hamra2013}). Estimating equations avoid these challenges through `stacking', where individual estimating functions are combined into a new overall estimating function. Stacking in \texttt{delicatessen} allows for the variance of $\hat{\delta}_{0.2}$ to be automatically and consistently estimated with the sandwich variance. The variance can then be used to construct Wald-type 95\% confidence intervals (CI) for each parameter. Letting $\psi_{EC}(\delta_{20}, \boldsymbol{\gamma})$ denote the estimating function for the 20\% effective concentration, the stacked estimating functions are
\[\psi(R_i, D_i, \boldsymbol{\theta}) =
\begin{bmatrix}
	\psi_{PL}(R_i, D_i; \boldsymbol{\gamma}) \\
	\psi_{EC}(\delta_{0.2}, \boldsymbol{\gamma})
\end{bmatrix}
\]
where $\boldsymbol{\theta} = (\boldsymbol{\gamma}, \delta_a)$. Using the built-in functions \texttt{ee\_loglogistic} and \texttt{ee\_loglogistic\_ed}, parameters in $\hat{\boldsymbol{\theta}}$ are estimated simultaneously. The estimated dose-response curve is shown in Figure 1b. The estimated 20\% effective concentration was 1.86 (95\% CI: 1.58, 2.14). 

Finally, we focus on standardization of results from one context to another. Consider the biomarker data in Kamat et al. (2012), which comes from a convenience sample of persons with HIV ($n=57$) \cite{Kamat2012}. When comparing the prevalence of recent drug use (cocaine or opiates) to a more representative cohort of persons with HIV in the United States \cite{D2021}, drug use was notably higher in the convenience sample (70\% versus 8\%). As there was differential biomarker expression by cocaine use, the differences in drug use patterns indicate that the biomarker expression results from Kamat et al. may not be generalizable. To account for differences in drug use, we standardized biomarker means from the convenience sample to secondary data representative of persons with HIV in the United States using inverse odds weights estimated from a logistic regression model \cite{Lever2016, Westreich2017}. Letting $\mu$ denote the mean for a biomarker and $\boldsymbol{\beta}$ denote the coefficients in the logistic model, the uncertainty for $\hat{\mu}$ depends on the variance of $\hat{\boldsymbol{\beta}}$, complicating variance estimation. Ignoring this dependence can lead to a false confidence in the precision of the biomarker results. To address this situation, the estimating functions for the logistic regression model and weighted mean are stacked together
\[\psi_w(S_i, X_i, \boldsymbol{\beta}) =
\begin{bmatrix}
	\left(S_i - \text{expit}((1, U_i) \boldsymbol{\beta}^T)\right) 1 \\
	\left(S_i - \text{expit}((1, U_i) \boldsymbol{\beta}^T)\right) X_i \\
	I(S_i=1) \times \frac{1 - \text{expit}(U_i^T \boldsymbol{\beta})}{\text{expit}(U_i^T \boldsymbol{\beta})} \times (\text{log}(B_i) - \mu)
\end{bmatrix}
\]
where the first two estimating functions are for the logistic regression model and the last is for the inverse odds weighted mean. Here, $U_i$ denotes drug use for individual $i$, $S_i$ denotes membership in the Kamat et al. study ($S_i=1$) or cohort ($S_i=0$), $\text{log}(B_i)$ is the log-transformed biomarker, and $\text{expit}(.)$ is the inverse logistic function. By stacking the estimating functions, uncertainty in the estimation of $\boldsymbol{\beta}$ is appropriately propagated into the variance estimate of $\hat{\mu}$. Results for select biomarkers are presented in Figure 1c. Notably, the means for log-transformed sIL-2R and IL-12 decreased after standardization. However, we note there are likely  important unmeasured differences between these studies, so these results should only be viewed as an illustration. While these highlighted examples involve straightforward dependencies, estimating equations and the sandwich variance estimator allow for dependence on multiple models and complex dependencies between parameters.

\section*{Acknowledgments}
The authors would like to thank Drs. Michael Love, Adaora Adimora, and the trainees on HIV/STI training grant at University of North Carolina at Chapel Hill for their feedback on earlier drafts of this work. Acknowledgments is to express our gratitude and does not imply their endorsement.

\noindent
The authors of this work were supported by grants from the National Institute of Allergy and Infectious Diseases (T32AI007001, K01AI177102, R01AI157758) at the time of the software development and writing of this paper. The content is solely the responsibility of the author and does not necessarily represent the official views of the National Institutes of Health.

\noindent
All source code is available on GitHub (\url{github.com/pzivich/Delicatessen}) and through the Python Package Index (PyPI). Documentation and further examples can be found on GitHub and the project website (\url{deli.readthedocs.io/en/latest/}). Feature requests, bug reports, or help requests can be done through the GitHub repository. 

\small
\bibliography{biblio}{}
\bibliographystyle{ieeetr}

%\end{multicols}

\newpage

\onecolumn

\section*{Appendix}

\subsection*{Empirical Sandwich Covariance Estimator}
The empirical sandwich variance estimator, $V(\hat{\boldsymbol{\theta}})$, for the asymptotic covariance matrix $V(\boldsymbol{\theta})$ is $B(\boldsymbol{\theta})^{-1} M(\boldsymbol{\theta}) \left(B(\boldsymbol{\theta})^{-1}\right)^T$. The `bread' matrix is defined as 
\[B(\boldsymbol{\theta}) = - E \left[ \nabla_{\boldsymbol{\theta}} \psi(\mathbf{Z}; \boldsymbol{\theta}) \right]\]
and $\nabla_{\boldsymbol{\theta}}$ denotes the vector of all partial derivatives of $\psi$ with respect to $\boldsymbol{\theta}$ and $E(\cdot)$ is the expected value function. The bread can be estimated by taking the sample average of the partial derivatives, either numerically approximated or computed exactly using automatic differentiation. The `meat' matrix is defined as the expected value of outer product, 
\[M(\boldsymbol{\theta}) = E \left[ \psi(\mathbf{Z}; \boldsymbol{\theta}) \cdot \psi(\mathbf{Z}; \boldsymbol{\theta})^T \right].\]
The meat can be estimated by taking the sample average of the outer product of the estimating functions. The estimated bread and meat matrices can then be composed into the empirical sandwich variance following the definition of $V(\boldsymbol{\theta})$. The finite sample variance estimator is then obtained by scaling the asymptotic variance estimate by $n$, i.e., $n^{-1} \times V(\hat{\boldsymbol{\theta}})$ \cite{Stefanski2002}.

\subsection*{Estimating Equations for the Log-Logistic Model}
The estimating equations for the 3-parameter log-logistic model with a lower dose-response limit of zero are
\[ \psi_{PL}(D_i, R_i, \boldsymbol{\gamma}) = \left[
\begin{matrix}
	2 \gamma_3 (Y_i - \hat{Y}_i)\frac{\gamma_2}{\gamma_1} \frac{\rho}{(1 + \rho)^2} \\
	2 \gamma_3 (Y_i - \hat{Y}_i) \log(D_i / \gamma_1) \frac{\rho}{(1 + \rho)^2} \\
	2(Y_i - \hat{Y}_i) \frac{\rho}{(1 + \rho)^2}
\end{matrix}
\right]
\]
where $\rho = (D_i / \gamma_1)^{\gamma_2}$ and $\hat{Y}_i = \theta_3 / 1+\rho$. The first, second, and third estimating equations are for $\gamma_1$, $\gamma_2$, and $\gamma_3$, respectively. The effective dose estimating equation is
\[\psi_{EC}(\delta_a, \boldsymbol{\gamma}) = \left[ \frac{\gamma_3}{\left(1 + \frac{\delta_a}{\gamma_1}\right)^{\gamma_2}} - \gamma_3(1-a) \right]\]
As previously stated, the effective dose estimating equation is a transformation of the 3-parameter log-logistic model parameters (i.e., it does not depend on $D_i$ or $R_i$).

\newpage

\onecolumn
\begin{landscape}

\begin{table}[h]
	\caption{Available features of open-source software implementing estimating equations\textsuperscript{*}}
	\centering
	\begin{tabular}{lcccccccc}
		\hline
		\multicolumn{1}{c}{}                   &          &                                                              & \multicolumn{6}{c}{Built-in Estimating Equations}                                                                                                                                                                                                    \\ \cline{4-9} 
		\multicolumn{1}{c}{Library}            & Language & \begin{tabular}[c]{@{}c@{}}User-Specified \\ EE\end{tabular} & \begin{tabular}[c]{@{}c@{}}Generalized \\ Linear Model \end{tabular} & \begin{tabular}[c]{@{}c@{}}Robust \\ regression\end{tabular} & \begin{tabular}[c]{@{}c@{}}4 parameter \\ log-logistic\end{tabular} & \begin{tabular}[c]{@{}c@{}}Causal Effect \\ Estimation \end{tabular}
		& \begin{tabular}[c]{@{}c@{}}Accelerated Failure \\ Time Models \end{tabular} 
		& \begin{tabular}[c]{@{}c@{}}Random-Effects \\ Meta-Analysis \end{tabular} \\ \hline
		\texttt{delicatessen} & Python   & x                                                            & x                                                                & x                                                            & x                                                                   & x             & x   & x    \\
		\texttt{statsmodels}  & Python   &                                                              & x                                                                & x                                                            &                                                                     &               &     &      \\
		\texttt{sklearn}      & Python   &                                                              &                                                                  & x                                                            &                                                                     &               &     &      \\
		\texttt{geex}         & R        & x                                                            &     x                                                            &                                                              &                                                                     &               &     &      \\
		\texttt{sandwich}     & R        & x                                                            &                                                                  & x                                                            &                                                                     &               &     &      \\
		\texttt{MASS}         & R        &                                                              & x                                                                & x                                                            &                                                                     &               &     &      \\
		\texttt{rlm}          & R        &                                                              &                                                                  & x                                                            &                                                                     &               &     &      \\
		\texttt{drc}          & R        &                                                              &                                                                  &                                                              & x                                                                   &               &     &      \\
		\texttt{Mestimation}  & Julia    & x                                                            &                                                                  &                                                              &                                                                     &               &     &      \\ \hline
	\end{tabular}
	\floatfoot{EE: estimating equation(s). IPW: inverse probability weighting, AIPW: augmented inverse probability weighting.\\
	* Available features based on most recent version of each software as of 2026/07/31.}
\end{table}

\end{landscape}

% \includepdf[pages=-, fitpaper=true]{appendix_code.pdf}

\end{document}